\documentclass[aip,floatfix,twocolumn,showkeys,citeautoscript,reprint,amsmath,amssymb,showpacs,epsf]{revtex4-1}
\usepackage{graphicx}
\usepackage{siunitx}
\usepackage{color}
\usepackage{changes}
\usepackage{epsfig}
\usepackage{tikz}

\begin{document}

\title{Electronic structure and transport properties of coupled CdS/ZnSe quantum dots}

\author{Simon Liebing}
\email{science@liebing.cc}
 \affiliation{Joint Institute for Nuclear Research Dubna, Bogoliubov Laboratory of Theoretical Physics, $141980$ Dubna, Russia}
 \affiliation{TU Bergakademie Freiberg, Institute of Theoretical Physics,  Leipziger Strasse 23, D-09596 Freiberg, Germany}
\author{Torsten Hahn}
 \affiliation{TU Bergakademie Freiberg, Institute of Theoretical Physics,  Leipziger Strasse 23, D-09596 Freiberg, Germany}
\author{Jens Kortus}
    \email{kortus@physik.tu-freiberg.de}
    \affiliation{TU Bergakademie Freiberg, Institute of Theoretical Physics,  Leipziger Strasse 23, D-09596 Freiberg, Germany}
\author{Bidisa Das}
\email{cambd@iacs.res.in}
\affiliation{Technical Research Center and School of Applied and Interdisciplinary Sciences, Indian Association for the Cultivation of Science, Jadavpur, Kolkata 700032, India}
\author{Arup Chakraborty} 
\affiliation{Department of Chemistry and Institute of Nanotechnology and Advanced Materials, Bar-Ilan University, Ramat Gan, 52900, Israel}
    \affiliation{School of Physical Sciences, Indian Association for the Cultivation of Science, Jadavpur, Kolkata 700032, India}
\email{arupchakraborty719@gmail.com}
\author{Indra Dasgupta}
    \email{sspid@iacs.res.in}
    \affiliation{School of Physical Sciences, Indian Association for the Cultivation of Science, Jadavpur, Kolkata 700032, India}


\begin{abstract}
   Electronic structure and transport characteristics of coupled CdS and ZnSe quantum dots are studied using density functional theory and non equilibrium Greens function method respectively.
   Our investigations show that in these novel coupled dots, the frontier occupied and unoccupied molecular orbitals are spatially located in two different parts of the coupled dot, thereby indicating the possibility of asymmetry in electronic transport.
   We have calculated electronic transport through the coupled quantum dot by varying the coupling strength between the individual quantum dots in the limits of weak and strong coupling.
   Calculations reveal asymmetric current {\em vs} voltage curves in both the limits indicating the rectifying properties of the coupled quantum dots.
   Additionally we discuss the possibility to tune the switching behavior of the coupled dots by different gate geometries.
\end{abstract}

\date{\today}

\maketitle

\section{Introduction}
Semiconductor heterostructures are a class of promising materials that have already found wide scale applications in optoelectronic devices, high electron mobility transistors, solar cells and light emitting diodes\,\cite{Kroemer,hetero_Alferov}.
The functionality of such heterostructures can be further controlled at nanoscale, as the quantum confinement effect of the charge carriers in individual nanoscale structures is combined with the changes in electronic properties due to heterostucture formation. 
This may provide  additional functionalization possibilities of the heterostructures at nanoscale.
In fact it has been shown that semiconductor heterostructures at nanoscale offer tunability of the band gap due to band-offset engineering at the interface in a wide range due to the quantum confinement of electrons\,\cite{NG_adv,dalui_chemical_2016,NG_PRB}.
Semiconducting heterostructures can be either of type-I or type-II depending on the relative alignment of conduction and valence band edges of the parent materials that form the interface.
In a typical type-II case with staggered band alignment, the lowest energy states for electrons and holes are situated in different semiconductors; therefore, at the interfaces electrons and holes tend to stay spatially apart due to the energy gradient.
Thus, type-II alignment is beneficial for application in  photovoltaic devices, with excited electrons and holes located at two spatially different parts of the material, thereby reducing the chances of recombination\,\cite{ivanov_type-ii_2007}.
Such type-II alignment have been realized in various kind of heterostructues at nanoscale namely core-shell nanocrystals\,\cite{eychmuller_quantum_1993,tian_coupled_1996,dabbousi_cdsezns_1997,kim_typeii_2003,yang_enhanced_2003,wang_superparamagnetic_2004,valerini_temperature_2005,he_microwave_2008}, multicomponent heteronanorods\,\cite{huang2014facile} and coupled quantum dots (QDs) \,\cite{NG_adv,dalui_chemical_2016,tian_coupled_1996}.
In the case of coupled dots, there are two separate dots only interacting at an defined interface, where in case of core-shell structures one material is completely encapsulated by the second one.
Thereby, core-shell structures have the disadvantage that not both materials can be contacted.
Recently, it has been shown that coupled QDs can capture photons and can be integrated as an active component within quantum dot sensitized solar cells\,\cite{dalui_chemical_2016}.
In addition, such type-II heterostructures also offer the possibility of charge transfer very similar to molecular heterojunctions. 
Molecular heterojuctions\,\cite{aviram_molecular_1974,hahn_gate_2014}, where charge transfer takes place between two molecules behaving as a donor and an acceptor, were earlier proposed to be useful in molecular electronic devices such as in molecular diodes or molecular switches. 
However, their use in real applications is still limited due to stability issues.
Analogously, type-II heterostructures at nanoscale, in particular type-II coupled QDs, may provide an alternative route to design diode and transistor devices at nanoscale.

While reports on such diode or transistor devices using semiconductor heterostructures (p-n junction)\,\cite{FET_science,Nature_recti} are available, not much is known on the transport properties of coupled dots, and how the relative orientation of the dots might affect their electronic and transport properties.
One question we want to address: Will the asymmetric couplings of the coupled quantum dots be found in the transport characteristics? 

Dalui \emph{et al.}\,\cite{dalui_chemical_2016} reported on the formation of interfaces between a fixed sized ZnSe QDs (\SI{4.5}{nm}) 
and three different sized CdS QDs (\num{2.6}, \num{4.2} and \SI{5.6}{nm}). 
These dots have sizes which makes them too large to compute their transport characteristics directly using atomistic electronic structure methods. 
Specifically the number of gold atoms required to simulate leads for such large systems is not computationally feasible.
Therefore we choose to reduce the number of atoms by introducing a suitable model system.
In our study we have scaled down the size of the coupled dots so that each individual model dot contains only 24 atoms, to keep the electronic transport modelling computationally tractable. 
Please note, these model coupled QDs are not only dynamically stable but also display the typical type-II behaviour that is known from dots with larger sizes as investigated experimentally\,\cite{NG_adv,dalui_chemical_2016,NG_PRB}. 
In order to explore the electronic structure and investigate the effect of the coupling strength of the CdS/ZnSe QDs, we have performed electronic structure studies based on density functional theory (DFT).
In addition, we have calculated the electronic transport through the coupled QD in a two-probe set-up and calculated the current$-$voltage characteristics of model devices and correlated the transport properties with the coupling strength of the QDs.
The most interesting effects in coupled QDs arise in cases where the electronic states in both dots are equal or very close in energy.  
Typically the states in QD  devices can be tuned via careful application of gate voltages on each of the individual dots.
It will be interesting to see whether the effects predicted on model systems\,\cite{van_der_wiel_electron_2002} could be realized in experimentally synthesized CdS/ZnSe coupled QDs.
The remainder of the paper is organized as follows, we shall discuss our methodology in section II, followed by our results on electronic structure and transport calculations in section III. 
Finally conclusions are provided in section IV.

\section{Methodology}
We have employed  DFT based electronic structure methods to accurately model the electronic properties of the model coupled QDs,
employing generalized gradient approximation (GGA) with Perdew-Burke-Ernzerhof (PBE) functional\,\cite{perdew_generalized_1996}.
All structures were subjected to geometrical optimization till the forces on each atom were below \SI{0.05}{eV/\AA}.
The geometries were optimized using NRLMOL\,\cite{pederson_strategies_2000,pederson_variational_1990,pederson_pseudoenergies_1991,porezag_infrared_1996,kortus_magnetic_2000} with the density functional optimized DFO basis set\,\cite{porezag_development_1997}.
The Vienna {\it ab-initio} Simulation Package (VASP)\,\cite{vasp1,vasp2} was used in order to double check all results.
We have used $\Gamma$-centered single point k-mesh to map the Brillouin-zone of these confined dots. 
Both studies gave similar results for geometry and electronic structure.
Finally,  all the structures were also optimized using hybrid functionals B3LYP\,\cite{B3LYP1,B3LYP2,B3LYP3} and B3PW91\,\cite{B3LYP1,B3PW91} with 6-31G** basis set for all atoms except Cd (Lanl2DZ basis set and pseudopotential) as implemented in Gaussian09 software\,\cite{Gaussian09}. 
Stability of the structures was investigated by vibrational frequency calculations. 
No imaginary modes were found and therefore the coupled QDs are found to be vibrationally stable.  
The electronic transport calculations presented are based on the nonequilibrium Green's function (NEGF) method as implemented in the \textsc{GPAW} code\,\cite{chen_ab_2012,enkovaara_electronic_2010}.
The \textsc{GPAW} transport code uses the Green's function of the central region defined by 
\begin{align} G(E) = \left[E S - H_C-\Sigma_L(E)-\Sigma_R(E)\right]^{-1} \end{align} 
where $S$ and $H_C$ are the overlap and Hamiltonian matrix of the scattering region.  
$\Sigma_{L/R}$ are the respective self energies of the leads.  
After the Green's function $G$ is solved self-consistently the transmission function $T$ is calculated using the following expression,
\begin{align}
  T(E,V) = Tr\left[G(E)\Gamma_L(E)G(E)^\dagger \Gamma_R(E) \right]
\end{align}
with $\Gamma_{L/R}(E)=i\left(\Sigma_{L/R}(E)-\Sigma_{L/R}(E)^\dagger\right)$. 
Therefore, $T(E,V)$ gives the transmission probability of an electron having an energy $E$ under an applied bias (and gate) voltage $V$.
The transmission function needs to be recalculated for each applied voltage.
Further the current through the junction is obtained by
\begin{align}
  I(V) = \frac{2e^2}{h}\int^{\mu_R}_{\mu_L}T(E,V) dE
\end{align}
where the electronic chemical potentials $\mu_{L/R}$ are connected to the applied bias voltage via $V=(\mu_L-\mu_R)/e$ ($e$ elementary charge)\,\cite{meir_landauer_1992}.
The current is calculated by integrating the self-consistent transmission function within the bias-dependent energy window spanned by $\mu_{L/R}$.
In addition to transport studies using \textsc{GPAW} code we have also done the electronic transport using Quantumwise software\,\cite{ATK1,ATK2} which follows same working principles described above based on NEGF. 
The DFT implementation in Quantumwise uses numerical atomic basis set to solve the Kohn-Sham equations. 
Double zeta polarized basis set was used for all atoms except Au (single zeta polarized basis set was used, to save computational resources) with GGA-PBE exchange correlation functional. 

\section{Results \& Discussions}
\subsection{Electronic Structure of the Coupled QDs }

The coupled quantum dot,  CdS/ZnSe is modelled by conceptually fusing two isolated semiconducting QDs, CdS and ZnSe, each having a fullerene-like hollow-cage structure with 12 cations and 12 anions.
Similar kind of small magic sized clusters of Zn$_{12}$O$_{12}$,\,\cite{ZnO_cluster} Cd$_{12}$S$_{12}$,\,\cite{CdS_cluster} Zn$_{12}$S$_{12}$,\,\cite{ZnS_cluster} In$_{12}$As$_{12}$\,\cite{InAs_cluster} etc, were reported earlier.
The lowest-energy structure for the isolated Cd$_{12}$S$_{12}$ cluster was found to be composed of eight hexagonal rings and six four-membered rings, having overall T$_h$ symmetry. 
Our calculations using B3LYP reveal that the hexagons have alternating Cd$-$S bond-lengths of \num{2.50} and \SI{2.60}{\AA} respectively, while the Cd$-$S bond-lengths appearing in the four-membered rings are all \SI{2.60}{\AA}, which share the edges with the hexagons. 
For the Zn$_{12}$Se$_{12}$ cluster which is structurally very similar to the Cd$_{12}$S$_{12}$ cluster, the hexagonal rings have alternating Zn$-$Se bond-lengths of \num{2.35} and \SI{2.42}{\AA} respectively, while the bond-lengths in the four-membered rings \SI{2.42}{\AA}. 
The shorter bond-lengths encountered in ZnSe QD compared to CdS indicate greater ionic character of the former.

To construct the model coupled QDs, we have connected the two different semiconducting QDs, Cd$_{12}$S$_{12}$ and Zn$_{12}$Se$_{12}$ of approximately similar sizes in many different ways and relaxed the resulting coupled structures, to locate the minima on the potential energy surface.
As these QDs are composed of hexagonal rings and four-membered rings, they can be linked together in three different ways; (i) by aligning individual QDs via hexagonal faces, where six atom pairs can interact (ii) interfacial alignment through the four-membered rings of the individual QDs, with four interaction atom pairs, and (iii) via edge to edge interactions, where two atom pairs can interact. 
For the latter case three different configurations can be built where different intra-dot bonds are aligned next to each other.

\begin{figure}[h]
\centering
\includegraphics[width=\linewidth]{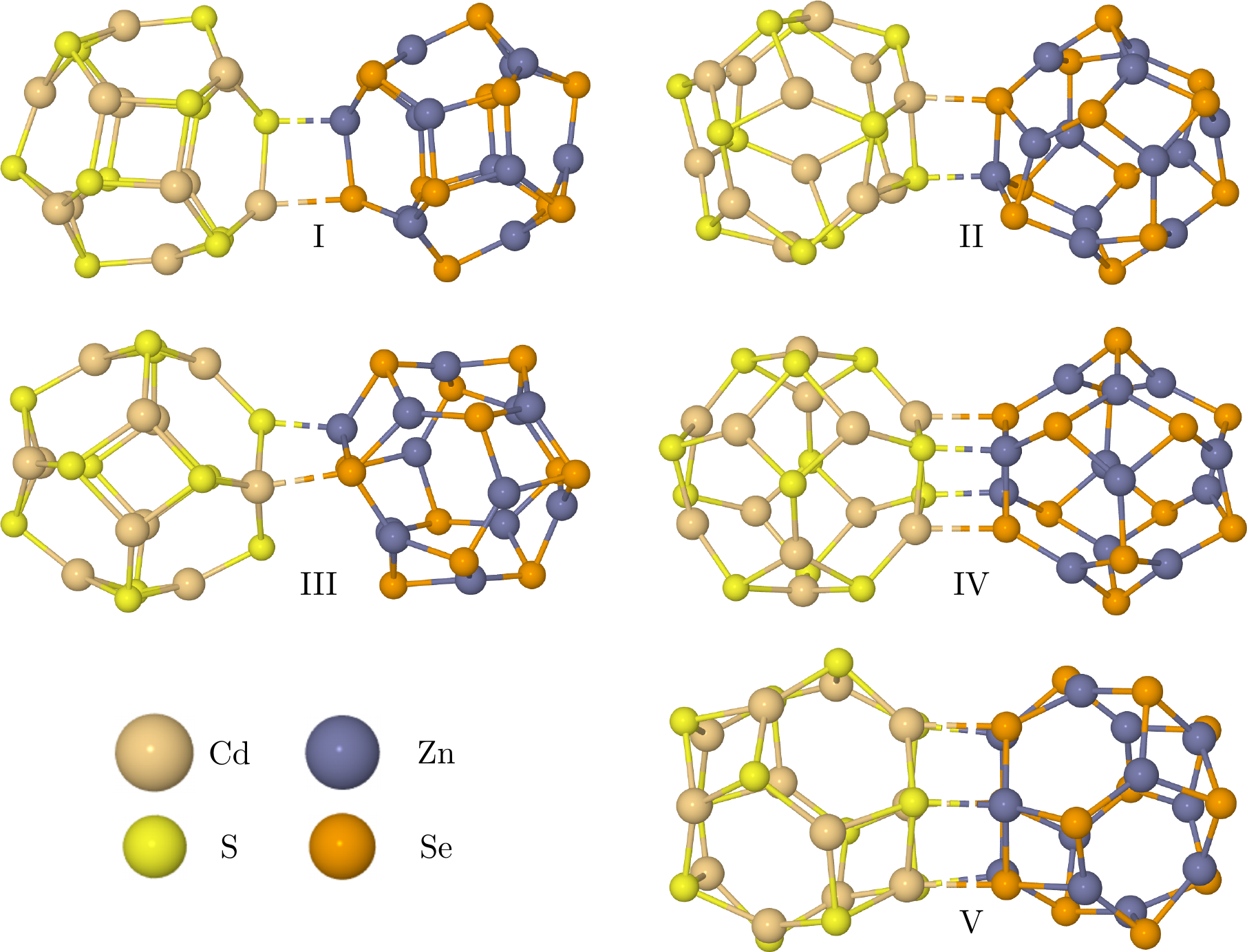}
\caption{Different configurations of CdS-ZnSe coupled dot. The different orientations of the dots towards one another changes the strength of possible interactions}
\label{structure}
\end{figure}

These  five different coupling possibilities are  shown in figure\,\ref{structure}. 
The different inter-cluster interactions are marked as configuration I - V throughout this paper. 
In detail, configuration I is formed by the edges between two hexagonal rings.  
The edge of a four-membered and a six-membered ring forms configuration II.
The edges of two hexagonal rings on one QD and the edge of a four-membered and a six-membered ring on the other QD is configuration III and four pair interactions between two four-membered rings (configuration IV) and six pair interactions between two hexagonal rings (configuration V) belonging to two different QDs  are also possible.

To investigate the stability of these coupled dots, we have calculated the respective interaction energies. 
The interaction energy ($\Delta$E) of coupled dots can be calculated from the total energies of coupled systems and their constituent dots. 
It is defined as follows.\,\cite{Zunger_delE}
\begin{eqnarray}
\Delta{E} = [E(\text{CdS}/\text{ZnSe}) - (E(\text{CdS}) + E(\text{ZnSe}))]
\end{eqnarray}
The calculated interaction energies (see Table-1) are found to be negative in all cases indicating that the structures of coupled dots have lower energy compared to the individual dots. 
The calculated interaction energy for the most stable configurations within PBE and B3LYP are \SI{-0.88}{eV} and \SI{-1.76}{eV} respectively. 
\begin{table}[h]
\caption{Interaction energy of different configurations for CdS/ZnSe QDs in eV with respect to both separated dots. 
Results from two different DFT methods are shown.}
\vspace{0.25in}
\centering
\begin{tabular}{c|rr}
 Configurations &  \multicolumn{2}{c}{$\Delta$ E in eV}  \\
& PBE & B3LYP \\ \hline 
I & -0.62 & -1.05 \\
II & -0.59 & -0.99 \\
III & -0.65 & -1.16 \\
IV & -0.74 & -1.41 \\
V & -0.88 & -1.76 \\
\end{tabular}
\label{Table_delE}
\end{table}

Above results show that the system with six pair interactions at the interface (configuration-V) is energetically favored.
Further for an easier comparison, we shall refer to configuration-I as weak-coupling and the configuration-V as strong-coupling, and all the other configurations II, III and IV will be referred to as intermediate coupling regime.
We start our discussion by considering the most stable case V, the strongly coupled CdS/ZnSe QD, and discuss its structural properties as calculated using B3LYP.

In configuration-V, the six-membered, hexagonal ring of Cd$_{12}$S$_{12}$ QD is interacting face$-$to$-$face with a hexagonal ring of Zn$_{12}$Se$_{12}$ QD on the other side, allowing interaction of six Zn-S and Cd-Se pairs. 
Strong interaction between the two individual clusters alter the inter atomic distances, not only in the inter-cluster region but also, in its near vicinity.
At the interface of the coupled QD, the hexagonal Cd{$_3$S$_3$ ring (connected to the Zn$_{12}$Se$_{12}$ QD) has alternating bond lengths which are $\sim$ \SI{0.1}{\AA} longer (\num{2.59} and \SI{2.72}{\AA}) than other rings and the hexagonal interfacial ZnSe ring also has longer alternating bond lengths (\num{2.43} and \SI{2.50}{\AA}). 
This shows that there is an overall increase in the ring size for both the dots due to strain, affecting the hexagonal rings at the interface. 
The increase in the coordination of atoms from three to four in the interface may also lead to longer bond-lengths. 
The Cd-Se and the Zn-S bridge bonds in the interfacial region are \SI{2.79}{\AA} and \SI{2.55}{\AA} respectively, longer than the usual bond-lengths calculated in the respective pure QDs. 

In contrast, the weakly coupled CdS/ZnSe dot in configuration I, show deviations of the bond distances from the uncoupled QDs only for the bridge-bonds and nowhere else.
The Cd-Se bridge bond in this case is \SI{2.91}{\AA} which is very long, while the Zn-S bridge bond is \SI{2.47}{\AA} which is slightly longer due to interfacial strain. 
This minimal effect on average bond distances for the coupled dot in configuration I also reflected in its reduced interaction energy compared to configuration V (Table\,\ref{Table_delE}). 
These trends in structural properties are also observed with the PBE calculations.
\begin{figure*}
\centering
\includegraphics[width=0.9\linewidth]{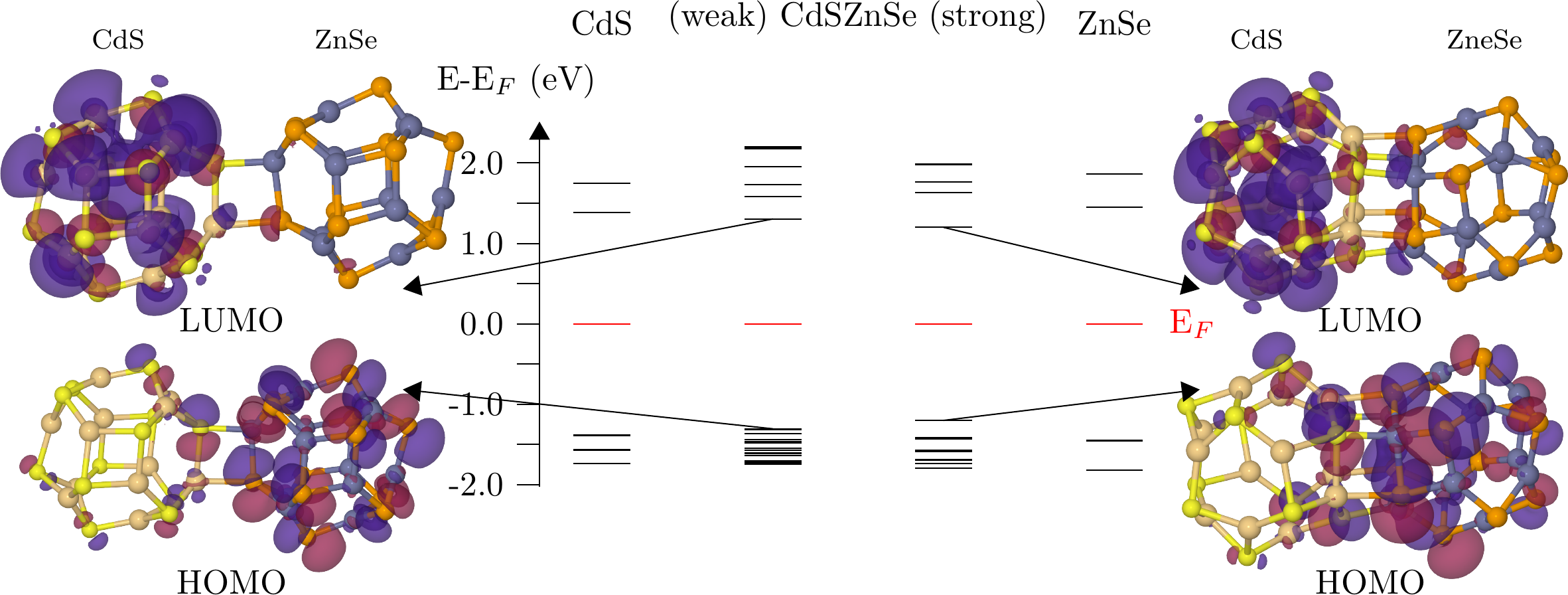}
\caption{The energy levels of individual dots and both the coupled dots and the isosurfaces of strong and weak coupling.}
\label{iso_level}
\end{figure*}

The  calculation of the band gaps of the isolated as well as the coupled QDs  are performed with care using different methods. 
As pure DFT functionals like PBE routinely underestimate the HOMO-LUMO gaps obtained from the energy difference of the highest occupied molecular orbital (HOMO) and lowest unoccupied molecular orbital (LUMO), the identification of HOMO-LUMO gaps ($\mathsf{E^{HL}_{G}}$) with the band gap is only one possible definition. 

Within DFT other common approach to obtain band gaps is the use of ionization potential (IP) and electron affinity (EA) for the definition of the band gap \(\mathsf{E^F_G} = \mathsf{IP} - \mathsf{EA}\).
The ionization potential is obtained from total energy differences of the positively charged system and the neutral one $\mathsf{IP}=\mathsf{E}(\mathsf{N}-1)-\mathsf{E}(\mathsf{N})$ and the electron affinity is defined as $\mathsf{EA}=\mathsf{E}(\mathsf{N}+1)-\mathsf{E}(\mathsf{N})$.
The $\mathsf{E^F_G}$ values are larger than the  $\mathsf{E^{HL}_G}$ ones, and present a better estimate for the real band gap.
The IP and EA obtained using NRLMOL with PBE and the corresponding energy gap value $\mathsf{E^{F}_{G}}$ together with the energy gap $\mathsf{E^{HL}_{G}}$ obtained from the difference of the energies of the HOMO and the LUMO for individual QDs  and coupled QDs using PBE and B3LYP  are displayed in Table\,\ref{tab:electronic} for comparison. 

It can be seen that the trend of the band gap for the individual QDs and coupled QDs are same in all the methods. 
The band gaps of the uncoupled CdS and ZnSe QDs are larger compared to their coupled counterparts both in the weak and the strong coupling regimes, suggesting type-II band alignment at the interface. 
\begin{table}[]
 \caption{Electronic properties of the single and coupled QDs  with
  $\mathsf{IP}$ ionization potential, $\mathsf{EA}$ electron affinity, $\mathsf{E^{F}_{G}}$ gap
  from the difference between IP and EA, $\mathsf{E^{HL}_{G}}$ the HOMO-LUMO gap. HOMO-LUMO gaps obtained using various DFT methods are given. All values are given in eV.}
  \vspace{0.25in}
    \centering
\begin{tabular}{c|rrrrr}
        ~  & $\mathbf{IP}$ & $\mathbf{EA}$ & $\mathbf{E^{F}_{G}}$  & $\mathbf{E^{HL}_{G}}$ &$\mathbf{E^{HL}_{G}}$\\
& & & & (PBE)  & (B3LYP)\\
    \hline \\
    CdS     &  7.58 & 2.10 & 5.48 & 2.77 & 3.69 \\
    ZnSe    &  7.60 & 1.92 & 5.68 & 2.90 & 4.28 \\
    weak    &  7.14 & 2.38 & 4.76 & 2.61 & 3.52 \\
    strong  &  7.03 & 2.39 & 4.64 & 2.40 & 3.34 \\
\end{tabular}
    \label{tab:electronic}
\end{table}

\begin{figure}
\centering
\includegraphics[width=0.8\linewidth]{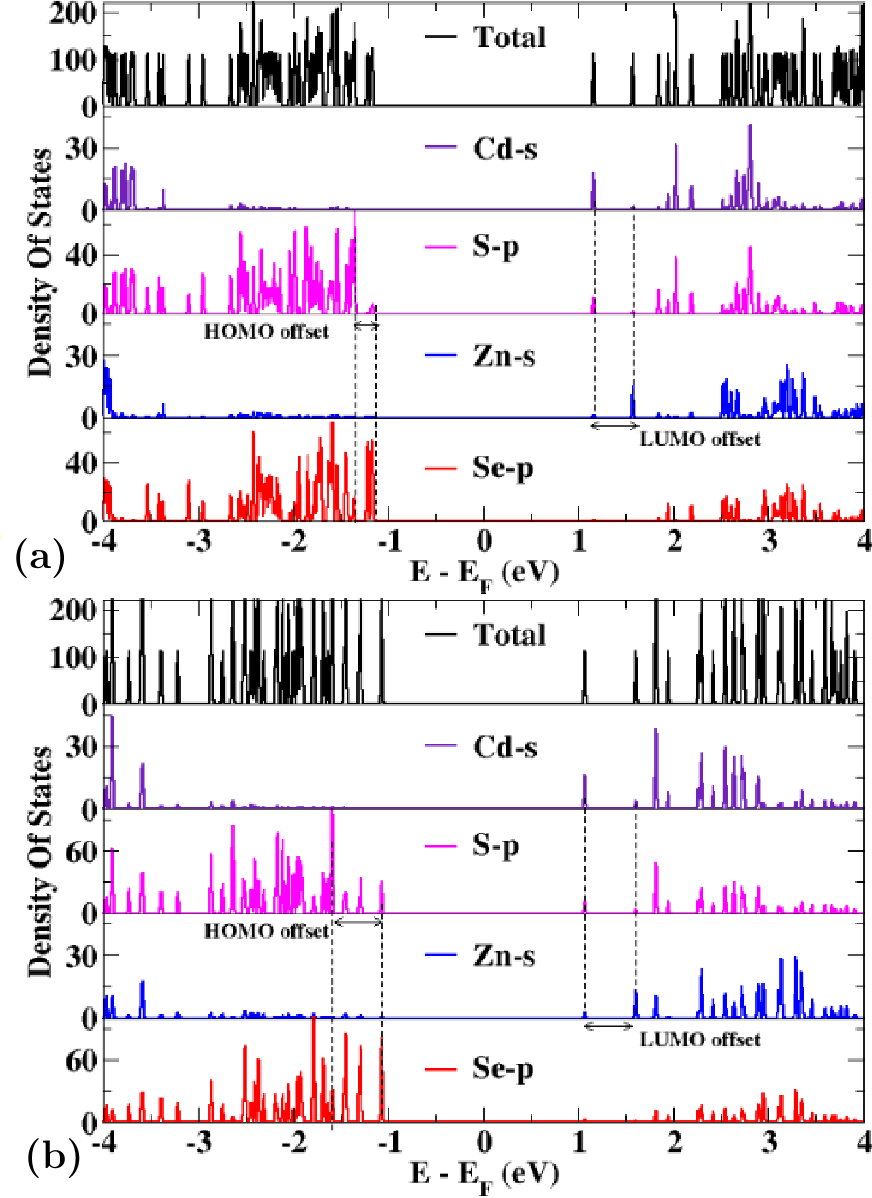}
 \caption{(a) Partial density of states for weak-coupling.(b) Partial density of states for strong-coupling.}
\label{dos_par}
\end{figure}
Next we discuss the energy levels of the isolated QDs as well as of the CdS-ZnSe coupled dot both in the weak and strong coupling regime using the PBE functional (see  Figure\,\ref{iso_level}). One can see that the energy levels are much closely spaced in the strong coupling regime in comparison to the weak coupling case.
The alignment of the states near the Fermi energy is expected to play an important role in determining its transport characteristics (see below). 

Additionally we have plotted the iso-surface of orbital density for HOMO and LUMO states as shown in Figure\,\ref{iso_level}.
The HOMO and LUMO states are localized at the ZnSe QD portion and CdS QD portion of the coupled QD respectively consistent with the expectation for a type-II heterostructure.
Further, one can see that the HOMO and LUMO of the strongly coupled QD  system extend more into the other QD due to the stronger coupling.

To obtain further insights on the nature of the level alignment and estimate the level offsets for the HOMO and the LUMO states at the interface we have plotted the partial DOS for the anion p and cation s states (see figure\,\ref{BD2}).
The partial DOS indicates that the occupied states near the gap are predominantly anion $p$-like whereas the unoccupied states near the gap mainly show cation $s$-character.
Interestingly the band offsets between the $p$-states of S and Se and the $s$-states of Cd and Zn are in accordance  with offsets typical of a type-II interface. Band-offsets change depending on the coupling (see figure\,\ref{dos_par}) and the different energy level alignments are expected to affect the transport properties which we shall investigate in the next section. 

\subsection{Models for two probe junctions with gold contacts to coupled dots}

The geometry of the two-probe setup including the gold Au(111) contacts for the weak and strong coupled dots are shown in Figure\,\ref{fig:scatt?model}.
\begin{figure}
\centering
 \begin{tikzpicture}
    \node at (0, 2.3) {\includegraphics[scale=0.09,origin=c]{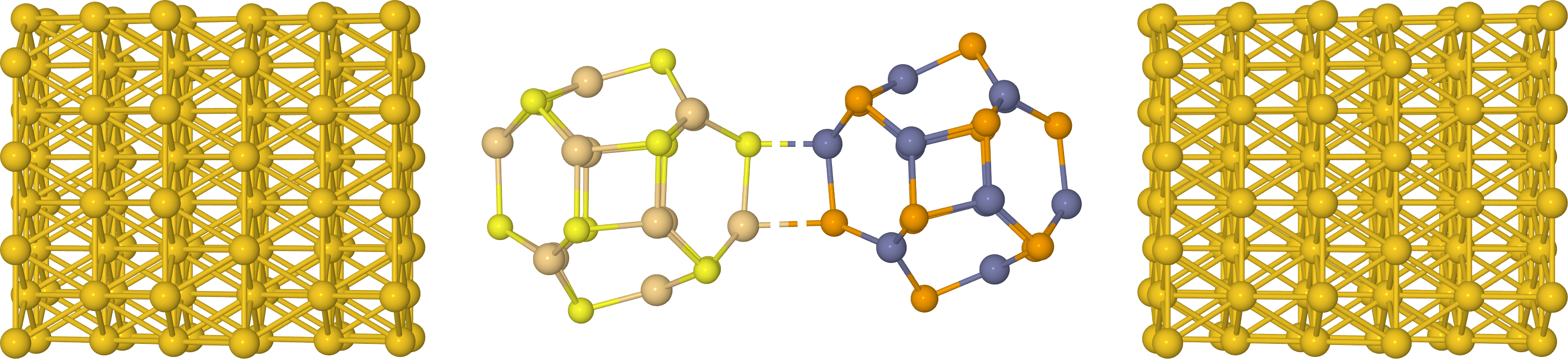}};
    \node at (0, 0) {\includegraphics[scale=0.09,origin=c]{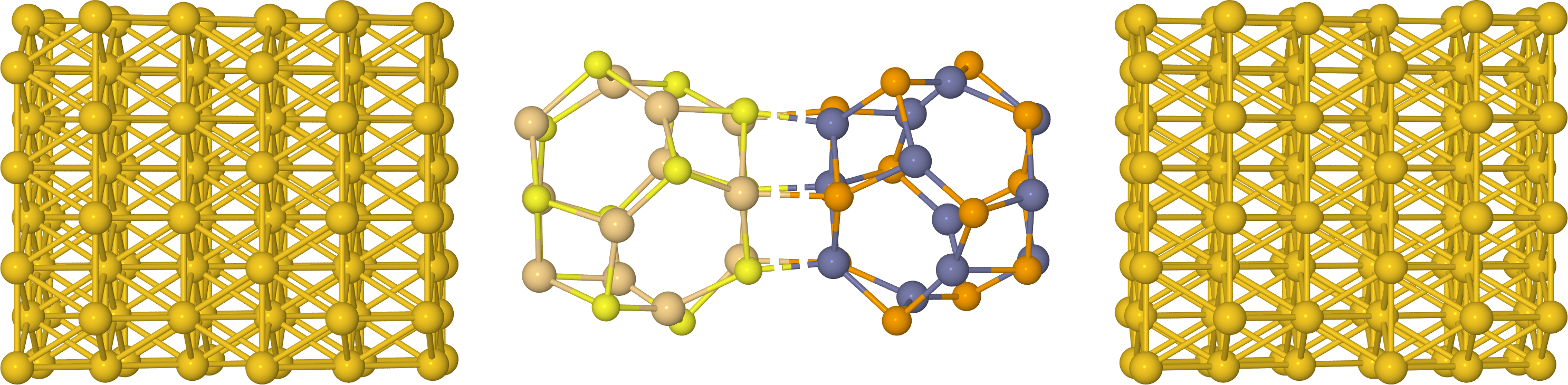}};
    \node at (-1.0,-1.3){CdS};
    \node at (1.1,-1.3){ZnSe};
    \node at (3.52,-1.3){Au (111)};
    \node at (-3.25,-1.3){Au (111)};
    \node at (0.1,1.1){strong};
    \node at (0.1,3.4){weak};
    \draw[color=black,  very thick,dotted](3.4,-1.9)--(3.4,3.5);
    \draw[color=black,  very thick,dotted](-3.4,-1.9)--(-3.4,3.5);
    \node at (3.85,-1.9){$\Sigma_R$};
    \node at (-3.85,-1.9){$\Sigma_L$};
     \node at (0.0,-1.9){scattering region};
\end{tikzpicture}
    \caption{Model of the weak\,(top) and strong\,(bottom) coupled CdS/ZnSe QD  between gold leads.
    On the right side is ZnSe\,(grey/orange) and on the left one CdS\,(beige/yellow).
    Additional one can see the regions of self energy and scattering region as used in the transport calculation.}
   \label{fig:scatt?model}
\end{figure}
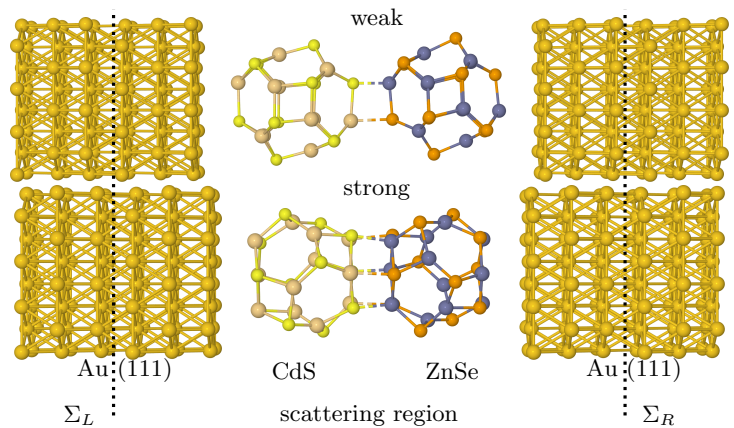
In order to build appropriate contact models and to avoid excessive computational requirement six layers of gold for each contact were used to model metallic electrodes.
The inner three gold layers are part of the scattering region and the outer three layers are repeated to represent metallic leads (see Figure\,\ref{fig:scatt?model}).

To find the minimum energy structure of the model devices we first attached only one gold contact to the respective coupled dot system and varied only the dot - lead interval to find the equilibrium distance.
In the second step the opposite side gold layer was placed by using the distance obtained in step one.
Finally we carried out a geometry optimization step for the whole system where the topmost gold layers together with the attached dots were allowed to relax. 
We double-checked all our transport calculations by using equivalent model systems within the electronic transport code implemented in the freely available \textsc{GPAW}\,\cite{enkovaara_electronic_2010} code and with the Quantumwise software\,\cite{ATK1,ATK2}.  

\subsection{Electronic transport properties of the coupled QDs }
The electronic transmission spectra calculated at zero bias for the two model junctions are shown in Figure\,\ref{BD2}a and \ref{BD3}a respectively. 
The transmission spectrum, while directly related to the current, also contains important microscopic information on the nature of transport channels and their coupling to the metallic electrodes. 

In general, the electronic states of the scatterer, which can conduct owing to their strong conjugation and overlap with the electrodes produce peaks in the transmission spectrum. 
The transmission eigenvalues for the given two-probe setup at a given energy and applied bias voltage are then obtained by diagonalizing the transmission matrix which are plotted in \ref{BD2}b and \ref{BD3}b respectively. 
The transmission eigenstate corresponds to the scattering state from the incoming left electrode to the outgoing right electrode. 
While the number of eigenvalues  indicate the number of individual transmission channels through the coupled QD, only the eigenstates with significant eigenvalues are plotted in Figure\,\ref{BD2}b and \ref{BD3}\,b. 

If several transmission channels are available at a particular energy, their sum and hence the transmission coefficient at this energy, may however be larger than 1 as is seen in the transmission spectra presented in Figure\,\ref{BD2}\,a and \ref{BD3}\,a. 
Higher values of transmission coefficients ($ >1 $) indicate parallel conducting pathways and the total transmission probability $T(E,V)$ can be defined as $T(E,V)=\sum_{n} T_{n}(E,V)$ where $n$ is the number of channels.
\begin{figure}
\centering
\includegraphics[width=1.40\linewidth]{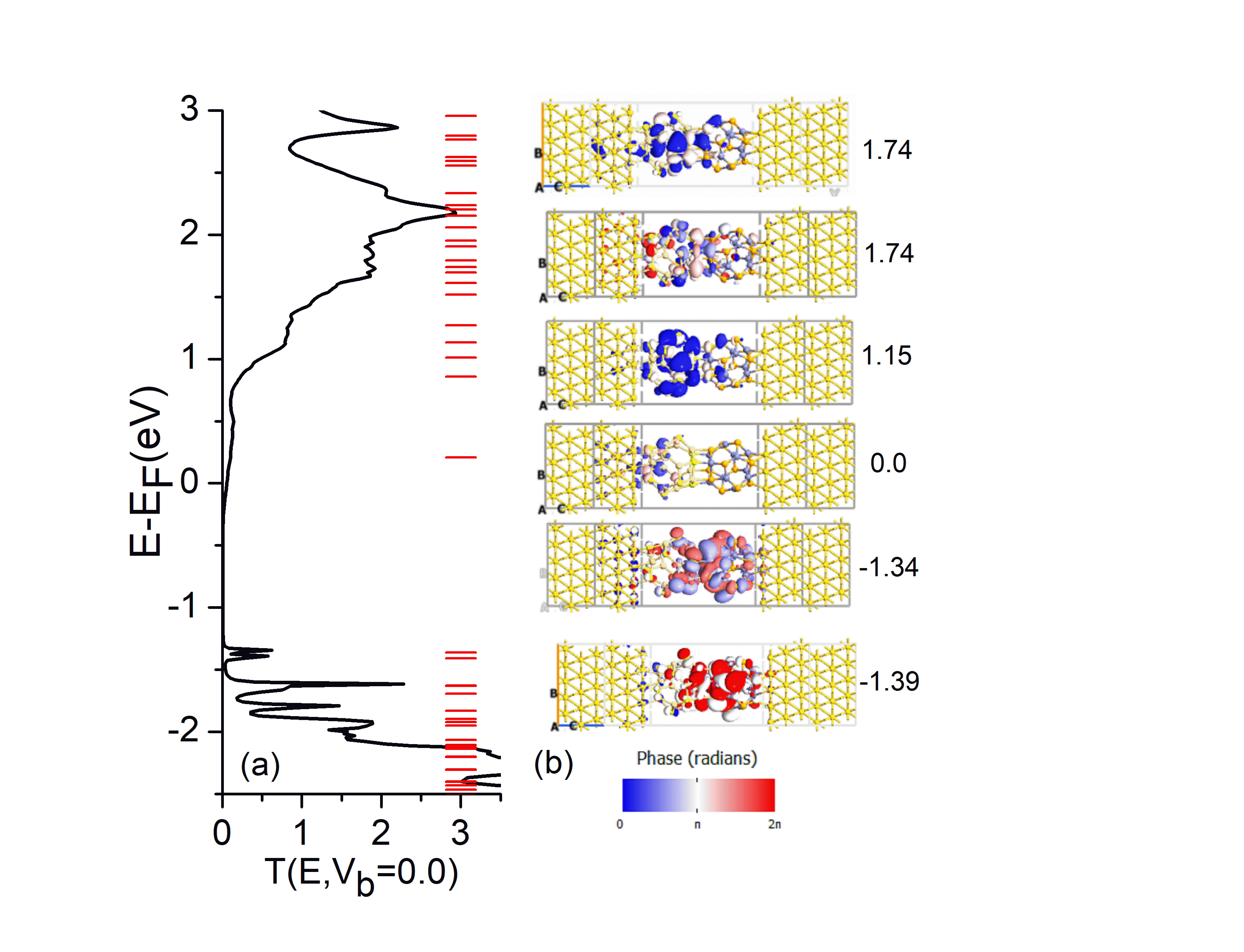}
\caption{\textbf{(a)} Transmission spectra presented for Au(III)-CdS/ZnSe QD -Au(III) two-probe setup considering both spins (strongly coupled case, V). The transmission spectra are calculated for zero applied bias. Horizontal lines shown at the right of each plot show states of the coupled dot in the two probe set up. These represent the states of the coupled dot modified by the presence of the electrodes (MPSH).  \textbf{(b)} The transmission eigenstates/channels calculated from the transmission spectrum T(E,V$_b$=\SI{0.0}{V}). The phases of the eigenstates are shown in two color format.  
}
\label{BD2}
\end{figure}

\begin{figure}
\centering
\includegraphics[width=1.40\linewidth]{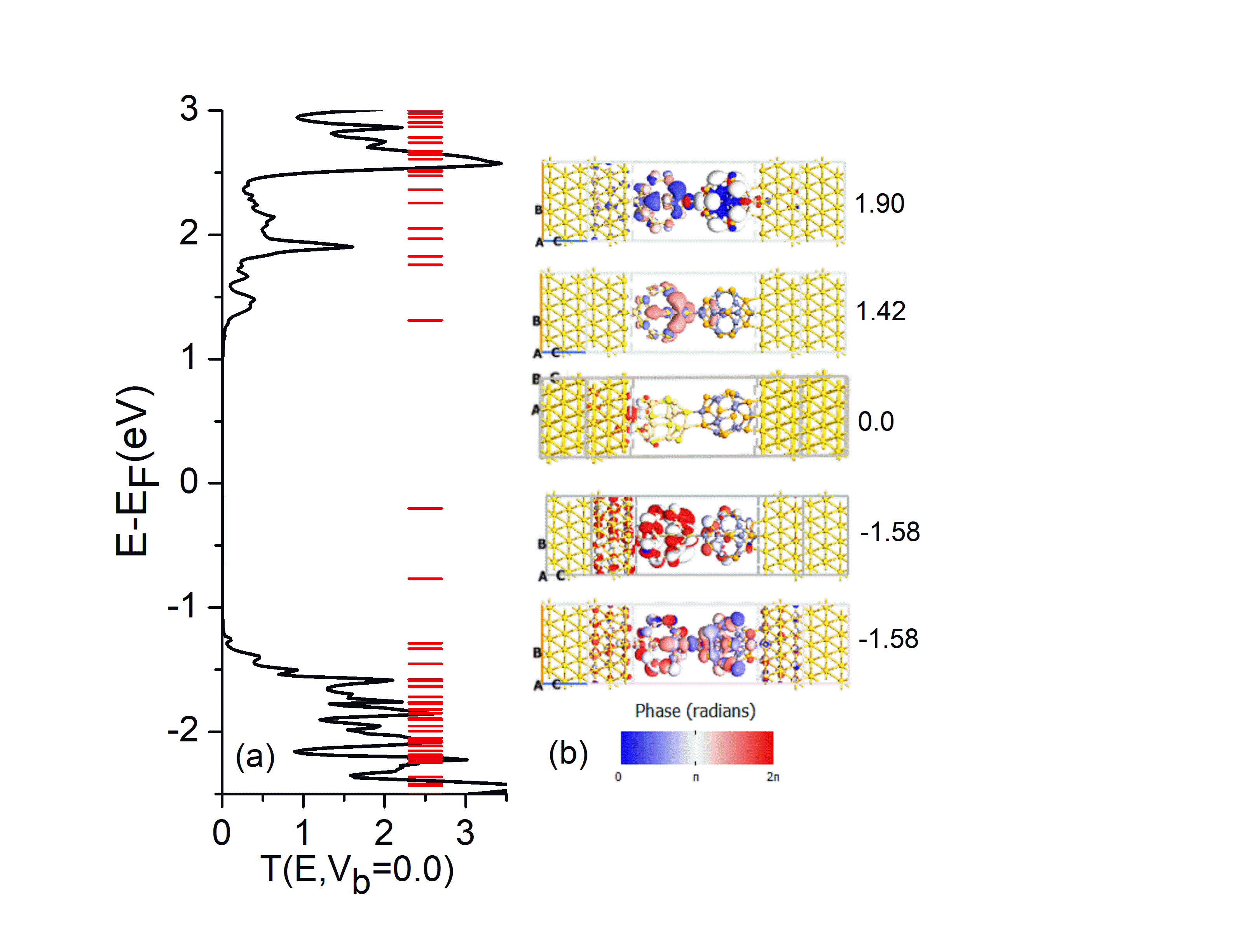}
\caption{\textbf{(a)} Transmission spectra presented for Au(III)-CdS/ZnSe QD -Au(III) two-probe setup considering both spins (weakly coupled case, I). The transmission spectra are calculated for zero applied bias. Horizontal lines shown at the right of each plot show states of the coupled dot in the two probe set up. These represent the states of the coupled dot modified by the presence of the electrodes (MPSH). \textbf{(b)} The transmission eigenstates/channels calculated from the transmission spectrum T(E,V$_b$=\SI{0.0}{V}). The phase of the eigenstates are shown in two color format.
 }
\label{BD3}
\end{figure}
There is a remarkable difference  between the transmission spectra of the strong and weakly coupled QDs in two-probe setups (Figure\,\ref{BD2}a and Figure\,\ref{BD3}a) in the gap region, near the Fermi energy. 
While the difference between the energy gaps of the weakly coupled and the strongly coupled isolated, contact-free coupled QDs is about \SI{0.2}{eV} (see Table II, $\mathbf{E^{F}_{G}}$), the difference in gap in the transmission spectrum (see Figure\,\ref{BD2}a and \ref{BD3}a ) in the two-probe setup is  approximately \SI{1.0} {eV}. 
In view of this the weakly coupled QD exhibits significantly lower conductance compared to the strongly coupled QD as obtained from transmission spectra at zero applied bias. 
Clearly the energy gap of the strongly coupled QD is much more affected by attaching the QD to the contacts than the weakly coupled QD. 
The fact is underlined by an analysis of the states that contribute to the respective transmission spectra.   
\begin{figure}[h]
 \centering
\includegraphics[width=1.15\linewidth]{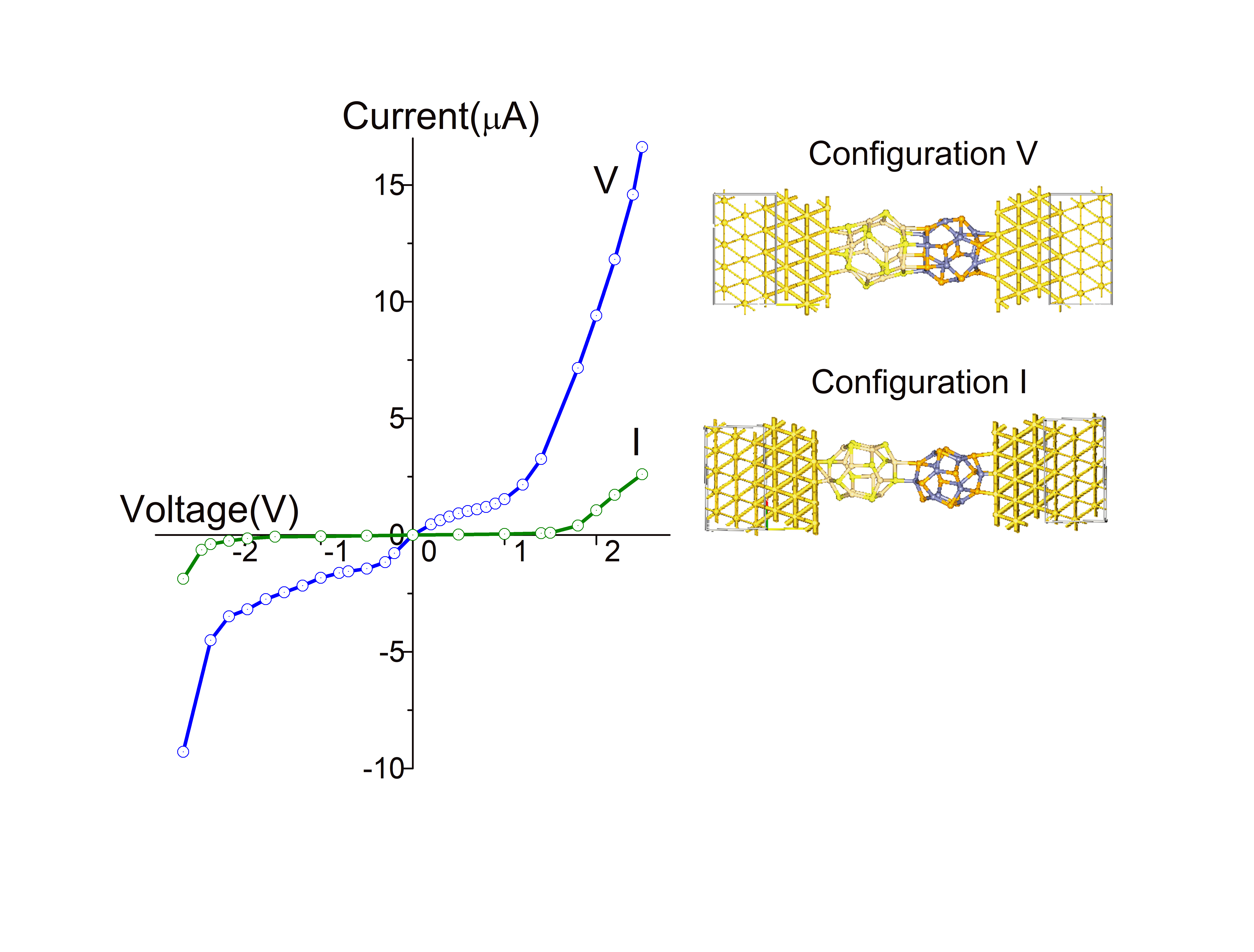}
 \caption{The calculated current versus voltage curve \textbf{I} for weakly coupled QD  and \textbf{V} for strongly coupled QD are shown in Au based two-probe setups which are shown at the right.}
\label{BD4}
\end{figure}

The HOMO level for the strongly coupled QD without contacts is found to be nearly degenerate as observed in Figure\,\ref{dos_par}a and this behavior is conserved in the two probe setup. 
The transmission eigenstates plotted in the Figure\,\ref{BD2}b  shows more delocalization of the coupled QD states at the interface of the strongly coupled QD along with significant contributions from the left electrode region corresponding to the incoming scattering state. 
Additionally, the eigenstate plotted at \SI{-1.34}{eV} also shows slight contribution of the right electrode indicating good conduction channel formed through the scatterer and the values of transmission coefficient are seen in the dual peaks around \SI{-1.4}{eV} in Figure\,\ref{BD2}a.
The inner orbitals which give higher $T(E,V)$ values are more delocalized (states not shown in the Figure). 

As discussed in the previous section, the HOMO is mainly located at the ZnSe QD, but the presence of the gold-contacts lead to a delocalization of the HOMO over both QDs in the coupled system to a certain extent forming a better transport junction.
The LUMO level of the strongly coupled QD is still localized within the CdS QD; it is significantly lowered in energy and is now found near the Fermi energy and the eigenstate is plotted in (Figure\,\ref{BD2}b). 
The eigenstate is found to be localized on CdS QD with contributions from the incoming left scattering state and shows a small transmission-coefficient.
The calculated conductance of the strongly coupled QD is \SI{2.16}{\micro S}.
The more conducting transmission eigenstates are plotted at 1.15 eV and two degenerate eigenstates at \SI{1.74}{eV} are also shown in Figure\,\ref{BD2}b. 
These eigenstates show significant contributions from the coupling region of the sandwiched QD which is reflected in their transmission co-efficient being $~$\num{2} in the plotted $T(E,V)$ Figure\,\ref{BD2}a.

In the  weakly coupled QD, the states between \num{-1} to \SI{1}{eV} of energy shown by red horizontal lines on the transmission spectra Figure\,\ref{BD3}a are localized states, and are not good conducting channels. 
The calculated eigenstates in this region thus show negligible eigenvalues and the transmission coefficient is almost zero showing no conduction. An inner state at \SI{-1.58}{eV} is seen to be delocalized, which shows two non-negligible eigenvalues corresponding to two simultaneous transmission eigenstates as shown in Figure\,\ref{BD3}b. 
In this setup the coupling to the contacts are lower and therefore there is small contribution to the transmission spectra arising from the LUMO located at \SI{1.42}{eV} and the corresponding transmission eigenstate (Figure\,\ref{BD3}b) is mostly localized on ZnSe QD. 
We also show another delocalized transmission eigenstate, at \SI{1.90}{eV}, responsible for the strong peaks in $T(E,V)$ (Figure\,\ref{BD3}b). 
The calculated conductance in this configuration is \SI{0.016}{\micro S}, which is almost two orders of magnitude smaller than the strongly coupled QD. 

The transmission spectrum calculated under zero bias voltage is not sufficient to describe the transport properties of a device. 
It is necessary to investigate the current versus voltage (I\mbox{-}V) curve, which can quantitatively describe the electron transport properties under finite bias voltages. 
The complete current\mbox{-}voltage, (I\mbox{-}V) curve of the model junctions in the bias region of [\SI{-2.0}{V}, \SI{2.0}{V}] are shown in Figure\,\ref{BD4}. 
First it is obvious that the conductivities of the two model systems differ significantly by almost two orders of magnitude (see also Figure\,\ref{fig:transport_model_bbias_log}). 
Additionally the I-V curves are asymmetric with respect to changing the sign of the bias voltage, in case of configuration I more than V. 
The latter gives rise to some rectification behaviour of the model QD junctions. 
This asymmetry stems from the fact that the states close to electrode Fermi energy are localized on either sides of the QDs (see Figure\,\ref{iso_level}), and thus contribute differently to electronic transport. 
Similar criterion has been exploited to design molecular rectifiers earlier \cite{aviram_molecular_1974,hahn_gate_2014} in donor-acceptor type molecules. 

In order to better understand the rectification characteristics for the coupled QD device, we investigated the change of the transmission spectra under various applied bias voltages.
This study shows that for both strongly coupled and weakly coupled dots with larger applied bias voltages, more states fall within the bias window, some of which are partially delocalized and thus act as electron transmission channels, especially for the strongly coupled QD and thus the total transmission probability and consequently the current increase. 
For weakly coupled QD the large energy gap in the $T(E,V)$ due to unavailability of suitable states is responsible for less increase in current within the applied voltage range studied.

Typically envisaged applications of coupled QDs  are the use of the respective setups as switching devices. 
Thus we investigated the I-V characteristics of the two coupled dot systems for different setups of gate electrodes.
Additionally it is possible to clarify the origin of the effects seen in the I\mbox{-}V characteristics. 
A gate setup where the gate voltage is applied to both dots simultaneously will change the energy levels of the coupled QD system relative to energy of the orbitals in the contacts.
Thus the main effect of such an setup is the change of QD and contact coupling.

\begin{figure}[h]
\centering
\includegraphics[width=0.98\linewidth]{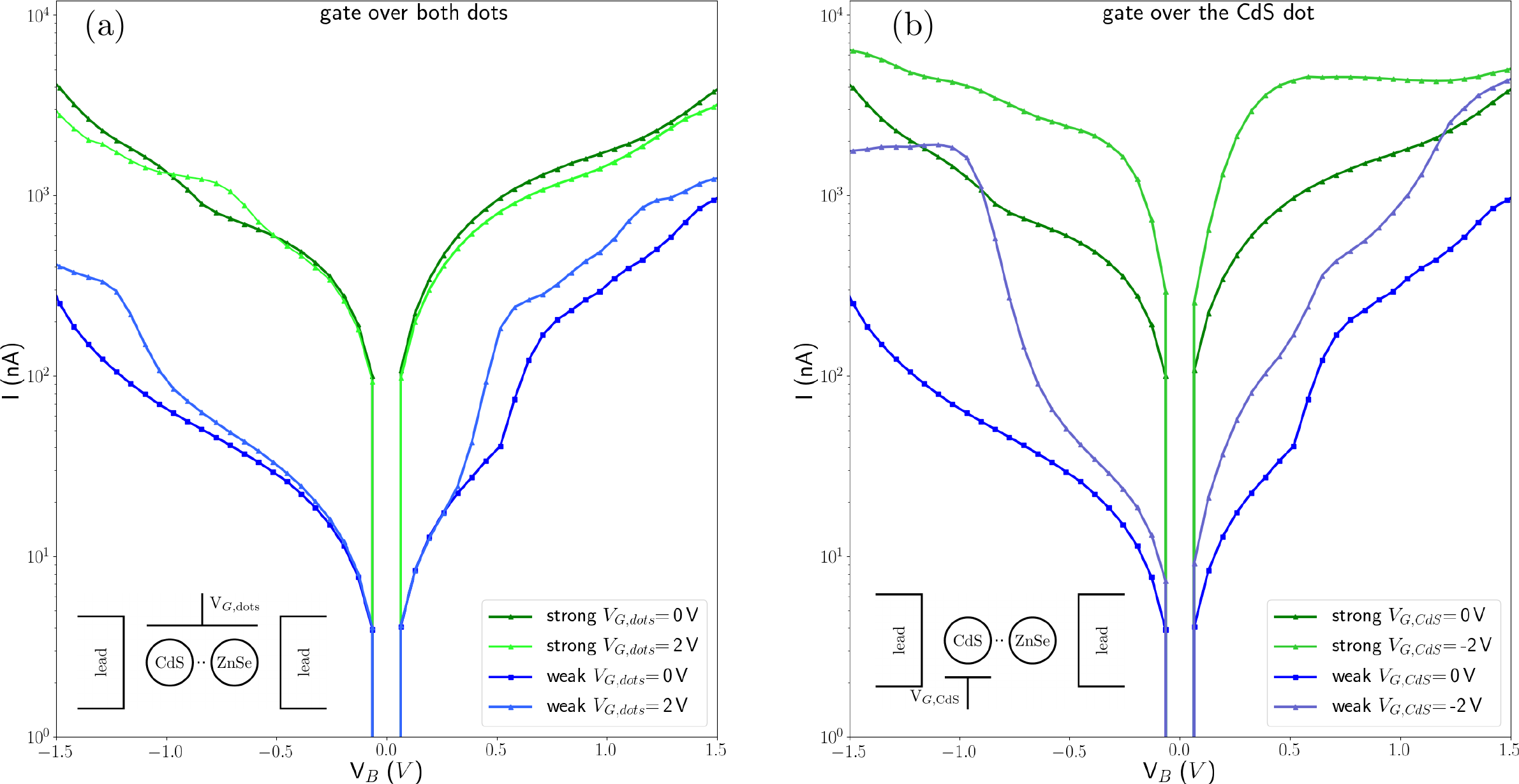}
\caption{\textbf{(a)} The current\mbox{-}voltage curves are corresponding to model device shown in Figure\,\ref{fig:scatt?model}. We plot the current on a logarithmic scale. In case of the weakly coupled QD  the current is about one order of magnitude smaller and slightly asymmetrical for the change of direction of the bias voltage. For the strong coupled dot the gate does not change much. In contrast to that for the weak coupling dot the current increases systematically and reaches nearly the strong coupling dots. \textbf{(b)} Gates over the CdS dot influence the current-voltage dependence. In this configuration there is an increase in the current for both couplings.
%
}
\label{fig:transport_model_bbias_log}
\end{figure}

A second setup where the gate is only applied to one dot of the coupled dot system is used to investigate the effect of shifting the energies of the levels in the CdS QD while keeping the ZnSe levels (almost) constant. 
In the latter setup the main effect is due to the change in coupling between the individual dots in the junction.  
The main results of this investigation together with schematic drawings of the respective junctions are shown in Figure\,\ref{fig:transport_model_bbias_log}.
Our study included gate voltages between \num{-8} and \SI{8}{V}. 
In general the effect of applying a gate voltage over both QDs in the coupled systems is small. 
The effect is even negligibly small for the strongly coupled QDs (see Figure \,\ref{fig:transport_model_bbias_log}a). 
On the other hand, the impact of applying the bias voltage on just the CdS QD is huge. 
As can be seen in Figure\,\ref{fig:transport_model_bbias_log}b, the current changes up to one order of magnitude upon application of the gate voltage. 
Further the asymmetry of the I\mbox{-}V curve of the weakly coupled QDs is amplified. 
These results can be interpreted in the following way.
Changing the coupling between the coupled QD systems and the gold contacts has very little effect on the I-V curves.

Thus the major effects that are responsible for the transport characteristics arise from the different electronic coupling between the individual QDs.
Qualitatively, a selected preparation of junctions with coupled dots in different bonding configuration would enable tuning of the junction properties.
Moreover the application of a gate voltage, especially for the weakly coupled configuration, would allow for the design of switching devices.  

\section{Conclusions}
We have studied the electronic structure and transport properties of CdS/ZnSe coupled QDs  using DFT and NEGF method respectively.
The electronic structure calculations reveal that different stable geometries are possible, which can be characterized by different interaction strengths. 
The significant feature of the electronic structure is the spatial separation of HOMO and LUMO on different QDs which are coupled together. 
This behavior can be found in all coupling configurations, which may be interesting and beneficial for photovoltaics application. 
The HOMO-LUMO photo excitation will result in charge separation across the coupled dots. 
Further, the use of QDs with variable size in principle offers the possibility to tune the band gap and therefore the absorption properties accordingly. 
The calculated I\mbox{-}V curves show, that the conductivity depends strongly on the interaction between the dots.
The conductivity of the strongly coupled dots is roughly two orders of magnitude larger compared with the weakly coupled dot.
In fact, the weak and strong coupling scenario provide limiting cases for the transport, for all investigated coupling scenarios. 
The usage of an additional gate voltage on one of the coupled dots allows to fine tune the coupling and thereby the transport behavior.
In particular it allows to switch the junction between conducting and non conducting mode effectively realizing an electronic switch.
The obtained I\mbox{-}V curves show strongly nonlinear behaviour, due to the asymmetry under bias inversion and diode-like rectification behaviour can be achieved. 
Further combinations of dots with largely different electron affinities / ionization potentials would enable the creation of intrinsic p-n junctions using coupled dots without doping.
Coupled dot systems are already under experimental investigation, and they are integrated as active component within QD sensitized solar cells. \cite{dalui_chemical_2016}
We hope that our results on transport properties for model coupled dot systems will motivate experimental research on such systems which may be interesting for electronic applications similar to the field of molecular electronics.

\section*{Acknowledgements}
Authors  would like to thank the DAAD (Germany) and DST (India) for making this work possible (PPP-India: "Electronic Structure and Transport in functional Nano Materials").
S.L. is financially supported by the European Union (European Social Fund) and by the Saxonian Government (grant no.101231954).
S.L., T.H. and J.K. want to thank the ZIH Dresden for providing computational resources to perform this research.
I.D. and B.D. thanks TRC (DST) for computational support.
\section*{References}
\bibliography{Coupled_SmallDot_ref}
\bibliographystyle{aipnum4-1}
\end{document}